\documentclass[twocolumn,prl]{revtex4-2}

\usepackage{graphicx}
\usepackage{booktabs}
\usepackage{amsmath}

\begin{document}

\title{Quantum theory is logically inevitable}

\author{Lars M. Johansen}
\affiliation{Department of Science and Industry Systems, University of South-Eastern Norway,Hasbergs vei 36, Kongsberg, N-3616, Norway}

\email{E-mail: lars.m.johansen@gmail.com}

\begin{abstract}
General relativity required the abandonment of Euclidean geometry. Here we show that quantum theory requires the abandonment of classical logic. We show that the Hilbert space representation of quantum theory is logically inevitable. There are no fundamental principles of quantum theory. We find an inevitable generalization of classical logic. The conjunction can take negative values, and reduces to the classical conjunction for order invariant measurements. The expectation of the conjunction is a generalized joint probability that can take negative values. A commutative conjunction leads to the Hilbert space formalism of quantum theory. Quantum theory applies both to microscopic and macroscopic systems. Classicality is represented by non-negativity of the generalized joint probability. We illustrate this by applying the logic to explain puzzling results in quantum cognition.
\end{abstract}

\maketitle

According to his research assistant Ernst Strauss  \cite{seeligHelleZeitDunkle1956}, Albert Einstein once stated that ``what really interests me is whether God could have created the world any differently; in other words, whether the demand for logical simplicity leaves any freedom at all''.

Quantum theory is probably the greatest intellectual achievement of the past century. It seems to be a general law of nature with universal validity. There are no experiments that has  shown any deviation from what it predicts. Obviously, there is something very general in the foundations of the theory. However, we don't know what that is. The theory is founded on abstract Hilbert space axioms with no direct connection with reality 
\cite{vonneumannMathematicalFoundationsQuantum1955}. This has led to a number of conceptual problems, in particular the so-called measurement problem. Many interpretations of quantum theory have been suggested. But the opinion on interpretation is more divided than ever \cite{schlosshauerSnapshotFoundationalAttitudes2013,cabelloInterpretationsQuantumTheory2017}.

The axiomatic Hilbert space formulation of quantum theory was published by von Neumann in his legendary textbook on the mathematical foundations of quantum mechanics \cite{vonneumannMathematicalFoundationsQuantum1955}. But soon after, he wrote in a letter to Birkhoff that ``I do not believe absolutely in Hilbert space any more" \cite{birkhoffLatticesAppliedMathematics1961}. In a joint paper, they set out to ``discover what logical structure one may hope to find in physical theories which, like quantum mechanics, do not conform to classical logic'' \cite{birkhoffLogicQuantumMechanics1936}. The paper initiated a long lasting research program in quantum logic.

The starting point for Birkhoff and von Neumann was to build the logic of quantum theory on a specific mathematical structure. An extensive discussion over mathematical structures and axiomatic foundations of the logic itself ensued. It led, in the end, to a ``labyrinth'' of possible quantum logics \cite{hardegreeChartingLabyrinthQuantum1981}. Since the axiomatic foundations of the logic itself were perhaps even more abstract than the Hilbert space foundations of quantum theory, the program made little or no impact on mainstream physics \cite{foulisHalfCenturyQuantumLogic1999}. The problem, originally posed by Birkhoff and von Neumann, of the \emph{operational} significance of logical connectives, never found an adequate solution \cite{kochenReconstructionQuantumMechanics2015}. The program did not lead to the desired conceptual clarification of quantum theory \cite{chiribellaQuantumPrinciples2016}, and interest in the program has waned in recent years.

The search for the conceptual foundations of quantum theory has now turned towards reconstructing quantum theory from intuitively comprehensible principles. This has been inspired by the way Einstein constructed his theory of relativity. The approach was initiated by Hardy, who derived quantum theory from ``five reasonable axioms" \cite{hardyQuantumTheoryFive2001}. He considered the operational procedures of preparations, transformations and measurements. This work gave rise to a number of reconstruction efforts in a tradition now called generalised probabilistic theories. Many reconstruction schemes have been found, in particular on basis of informational principles (see e.g. \cite{chiribellaQuantumTheoryInformational2016}). However, a unique identification of fundamental principles seems far away \cite{fuchsNegativeRemarksOperational2016}.

George Boole, the founder of the mathematical formulation of classical logic, in his classic treatise  ``The Mathematical Analysis of Logic'' \cite{booleMathematicalAnalysisLogic2009} assumed that ``the result of two successive acts is unaffected by the order in which they are performed''. He represented this by a commutativity relation. Otherwise, he stated, ``the entire mechanism of reasoning, nay the very laws and constitution of the human intellect, would be vitally changed. A Logic might indeed exist, but it would no longer be the Logic we possess'' \cite{booleMathematicalAnalysisLogic2009}.

Boole was thinking about mental acts. We know today that mental acts are order dependent. For example, people may respond differently to surveys depending on the order of questions asked. This is a standard textbook material in research methods (see e.g.  \cite{cohenResearchMethodsEducation2017}). In recent years, a new research area related to order dependence in human cognition has developed (see e.g. \cite{busemeyerQuantumModelsCognition2012}). And order dependence is of course well known in quantum theory. For example, if the spin of an electron is measured sequentially along two different directions, the result will depend on the order of the operations \cite{susskindQuantumMechanicsTheoretical2015}.

\section*{The logic of ordered questions}

In this paper, when using the term ``quantum theory'' we will essentially be be referring to the axiomatic Hilbert-space formulation of the theory. This is in analogy with how it's done in the field of quantum information (for a more careful distinction, see \cite{mullerInformationTheoreticPostulatesQuantum2016}). We will not address unitary time development, only the time development associated with measurements. This is where the major conceptual problems of quantum theory appear.

Basic to our analysis is the notion of \emph{yes-no questions}. Any experiment may be formulated as a yes-no question. We may consider any event, as a subset of sample space, and ask if the event happened or not. If the answer to the question $A$ is ``yes'', we assign the value $A=1$. If the answer is ``no'' we assign the value $A=0$. Therefore, per definition, a yes-no question is idempotent, $A^2=A$. Associated with a question $A$ is also the complementary question $\bar{A}$. The answer to $\bar{A}$ is always the opposite to the answer to $A$. We write this symbolically as
\begin{equation}
    A + \bar{A} = I_A
\end{equation}
where the value of $I_A$ is 1.

Consider now a sequence of two yes-no questions, $A$ and then $B$. We call this a \emph{sequential yes-no question}, and represent it by the \emph{sequential conjunction} $A \sqcap B$. The answer to a sequential question is also always ``yes'' or ``no''. It is ``yes'' if the answer to both questions is ``yes''. Otherwise, if the answer to one or both of the questions is ``no'', the answer to the sequential question is also ``no''. Therefore, the value of $A \sqcap B$ is the value of $A$ multiplied with the value of $B$ obtained afterwards. In contrast to Boole's approach, we allow the product to be non-commutative, allowing the question order to be of significance.

We now form the \emph{exclusive disjunction}
\begin{align}
		A \oplus B &= A \sqcap \bar{B} + \bar{A} \sqcap B.
		\label{eq:XOR}
\end{align}
The combined sequential question $A \oplus B$ consists of two disjoint sequential questions $A \sqcap \bar{B}$ and $\bar{A} \sqcap B$. The answer to these two sequential questions can never be both ``yes'' at the same time. Therefore, we can write the disjunction as a sum.

We now consider exclusive disjunctions of both question orders, $A \oplus B$ and $B \oplus A$. We substitute for the complementary questions. Since the answer to a question and it's complementary questions contains the same information, the sequential conjunction can be treated as distributive in this case. We therefore have
\begin{align}
		A \oplus B - B \oplus A = 2  \left ( B \wedge A - A \wedge B \right )
		\label{eq:tautology}
\end{align}
where we have introduced the \emph{logical conjunction}
\begin{align}
		A \wedge B = A \sqcap B + \frac{1}{2} \left (  B - B_A \right ).
		\label{eq:logcon}
\end{align}
We write down connectives only for one question order. For simplicity, we have introduced the notation $B \sqcap I_A = B$ and $I_A \sqcap B = B_A$ for single questions. Here $B_A$ refers to the question $B$ after a nonselective question $A$, i.e. after disregarding the answer to the question $A$. It is easily shown that the logical conjunction satisfies the marginality relations
\begin{align}
\begin{split}
	A \wedge B + A \wedge \bar{B} &= A,\\
	A \wedge B + \bar{A} \wedge B &= B.
\end{split}
	\label{eq:logmarg}
\end{align}
The last term on the r.h.s of Eq. (\ref{eq:logcon}) is the non-classical correction to classical logic. This term is what differentiates the logic from a Boolean algebra. The possible values for the logical conjunction ${A \wedge B}$ for all possible answers to the questions $A$, $B$ and $B_A$ are given in table \ref{tab:con}. Notably, it may take negative values. If $B=B_A$ the logical conjunction is Boolean.

\begin{table}
\centering
        \caption{Value-table for logical connectives.}\label{tab:con}
        \vskip 0.3cm
        \begin{tabular}{lcccc}
        	\toprule
            \multicolumn{5}{c}{\textbf{Conjunction $A \wedge B$}} \\
            \toprule
            &\multicolumn{2}{c}{$B  = 0$}
            &\multicolumn{2}{c}{$B  = 1$} \\ 
            \cmidrule(r){2-3}\cmidrule(r){4-5}   
            &$B_A = 0$&$B_A = 1 $&$B_A = 0$&$B_A = 1$ \\
            \midrule
            $A = 0$ & 0 & $- \frac{1}{2}$ & $\frac{1}{2}$ & 0 \\
            $A = 1$ & 0 & $\frac{1}{2}$ & $\frac{1}{2}$ & 1 \\
             \toprule
            \multicolumn{5}{c}{\textbf{Inclusive disjunction $A \vee B$}} \\
	        \toprule
            &\multicolumn{2}{c}{$B = 0$}
            &\multicolumn{2}{c}{$B = 1$} \\ 
            \cmidrule(r){2-3}\cmidrule(r){4-5}   
            &$B_A = 0$&$B_A = 1 $&$B_A = 0$&$B_A = 1$ \\
            \midrule
            $A = 0$ & 0 & $\frac{1}{2}$ & $\frac{1}{2}$ & 1 \\
            $A = 1$ & 1 & $\frac{1}{2}$ & $\frac{3}{2}$ & 1 \\   
            \bottomrule
        \end{tabular} \\
\begin{flushleft}
The possible values of the logical connectives as a result of answers to the questions $A$, $B$ and $B_A$. Both connectives are four-valued. Boolean truth-values are reproduced when  $B_A = B$ (leftmost and rightmost columns).  Non-classical half-integer values appear when $B_A \ne B$. The conjunction can take a minimum value of $-1/2$, and the inclusive disjunction can take a maximum value of $3/2$.
\end{flushleft}
\end{table} 

By rearrangement and simplification of Eq. (\ref{eq:tautology}) we obtain equations of the form
\begin{align}
		A \oplus B +2  A \wedge B = A + B.
		\label{eq:loginv}
\end{align}
In analogy with Boolean algebra, we can also \emph{define} the inclusive disjunction as
\begin{align}
		A \vee B = A + B - A \wedge B.
		\label{eq:incdisdef}
\end{align}
From (\ref{eq:loginv}) we can then show that
\begin{align}
\begin{split}
		 A \vee B &=  A \oplus B + A \wedge B,
\end{split}
\end{align}
which also is in correspondence with classical logic.

The possible values for the inclusive disjunction $A {\vee} B$ for all possible answers to the questions $A$, $B$ and $B_A$ are given in table \ref{tab:con}. We note that it can exceed unity.  If $B=B_A$ the inclusive disjunction is also Boolean.

\section*{Logical joint probabilities}

The expected value of an question $A$ is the \emph{probability} that the question $A$ gives the answer $A=1$, $\langle A \rangle = P(A=1)$. ${\langle A  \sqcap B \rangle}$ is the sequential probability of first obtaining the answer $A=1$ and then $B=1$.  The \emph{logical joint probability} $\langle A \wedge B \rangle$ is the expected values of the conjunction (\ref{eq:logcon}).

The logical joint probability is a generalization of the real part of the Kirkwood-Dirac quasi-probability distribution \cite{kirkwoodQuantumStatisticsAlmost1933,diracAnalogyClassicalQuantum1945}. 
The logical conjunction (\ref{eq:logcon}) has been encountered before in the analysis of successive projective measurements in quantum mechanics \cite{johansenQuantumTheorySuccessive2007}. In this case, the logical joint probabilities are order invariant. However, in the papers \cite{johansenQuantumTheorySuccessive2007,johansenReconstructingWeakValues2007}, the evolution due to observation was represented by the state. Here, the evolution is represented by the questions. It has also been shown that this structure is related to weak values \cite{aharonovHowResultMeasurement1988b,johansenReconstructingWeakValues2007}.
The logical joint probability can take negative values \cite{johansenNonclassicalityWeakMeasurements2004,johansenNonclassicalPropertiesCoherent2004}.

Wigner demonstrated that quantum states can be represented as quasi-probabilities over phase space \cite{wignerQuantumCorrectionThermodynamic1932}. These quasi-probabilities can take negative values for some quantum states. Soon after, Kirkwood derived a quasi-probability over phase space that could take complex values \cite{kirkwoodQuantumStatisticsAlmost1933}. This was further generalized by Dirac to arbitrary pairs of observables \cite{diracAnalogyClassicalQuantum1945}. The Kirkwood-Dirac distribution is closely related to the concept of weak measurements and weak values \cite{aharonovHowResultMeasurement1988b}. The weak value of an observable may exceed the eigenvalue spectrum.  This requires that the real part of the Kirkwood-Dirac distribution is negative \cite{johansenNonclassicalityWeakMeasurements2004}.

\begin{figure}[h]
\begin{flushleft}
\caption{Order-invariant logical joint probabilities}\label{fig:ClintonGore}
\includegraphics[width=0.5\textwidth]{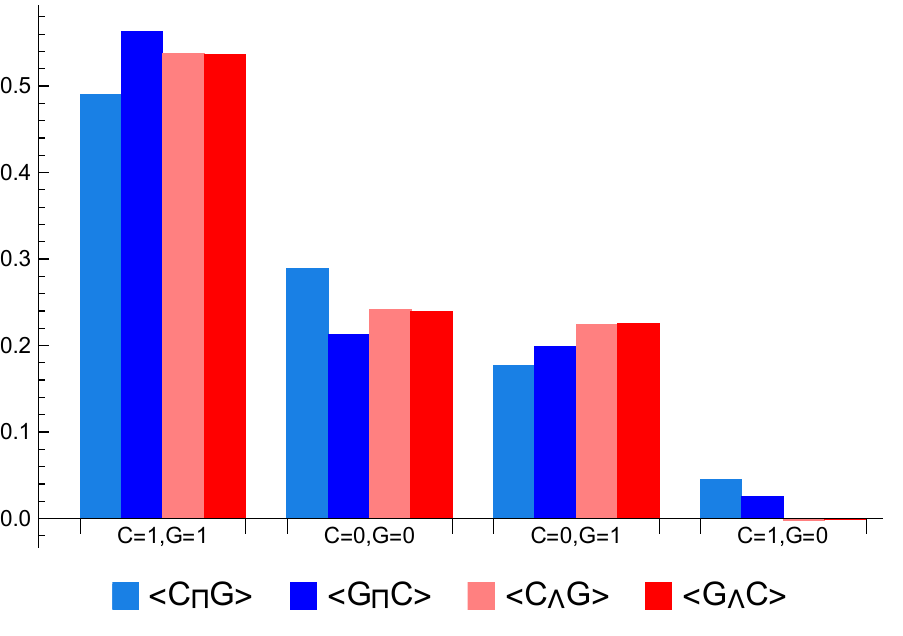}\\
In a Gallup poll conducted during September 6–7, 1997, people were asked "Do you generally think [Bill Clinton/Al Gore] is honest and trustworthy?" Half of the 1002  respondents were asked about Clinton first and Gore afterwards, the other half where asked the same questions in the opposite order. The results  exhibited a striking order effect. The data were examined in \cite{wangContextEffectsProduced2014} and found to satisfy a so-called quantum question equality \cite{wangQuantumQuestionOrder2013}. We have reconstructed the logical joint probabilities. Observed sequential probabilities are shown in blue hues, and reconstructed logical joint probabilities are shown in red hues. $C$  is the question if Clinton is honest and trustworthy, $G$ is the question if Gore is honest and trustworthy. The order effect on sequential probabilities is apparent. Nevertheless, we see that ${\langle C \wedge G \rangle}$ and ${\langle G \wedge C \rangle}$ are virtually identical. The observation is approximately ideal. The case $C=1,G=0$ is particularly striking. Here both sequential probabilities are positive whereas the logical joint probabilities are essentially zero. In fact, they are both slightly negative, but the effect is not statistically significant.
\end{flushleft}
\end{figure}

In Ref. \cite{wangQuantumQuestionOrder2013} a socalled ``quantum question equality'' was derived from quantum mechanical principles. In our notation, it reads $\langle A \oplus B \rangle = \langle B \oplus A \rangle$. This implies by (\ref{eq:loginv}) that $\langle A \wedge B \rangle = \langle B \wedge A \rangle$. In \cite{wangContextEffectsProduced2014} the quantum question equality was tested on a number of opinion polls. It was found that a requirement for the opinion polls to satisfy the quantum question equality was that questions should be asked successively with no additional information inserted in or between questions \cite{wangContextEffectsProduced2014}. An example of a poll satisfying the quantum question equality is shown in Fig. \ref{fig:ClintonGore}. We see that although the sequential probabilities are strongly order dependent, the calculated logical joint probabilities are nearly perfectly order invariant. We also note that all logical joint probabilities are non-negative, which is typical of the classical regime.

\section*{Commutative logic, Jordan algebras and Hilbert space}

Eq. (\ref{eq:tautology}) and the logical conjunction (\ref{eq:logcon}) apply to questions of any form, and regardless of anything that happens during or between the two questions. In general, none of the logical connectives are commutative. But of particular interest are commutative logical connectives. This would correspond to a particular subclass of sequential questions. Thus, we \emph{define} sequential questions as \emph{ideal} if the logical conjunction is commutative,
\begin{equation}
		 A \wedge B = B \wedge A.
		\label{eq:comcon}
\end{equation}
Because of Eqs. (\ref{eq:tautology}) and (\ref{eq:incdisdef}), this also implies that the exclusive and inclusive disjunction is commutative.

We conjecture that a commutative logic can be represented by an associative, non-commutative algebra. By making the L\"uders \emph{ansatz} \cite{luedersUeberZustandsaenderungDurch1950}
\begin{align}
		A \sqcap B \rightarrow A {\cdot} B {\cdot} A
		\label{eq:lueders}
\end{align}		
it follows that
\begin{align}
		B_A = A \sqcap B + \bar{A} \sqcap B \rightarrow A {\cdot} B {\cdot} A + \bar{A} {\cdot} B {\cdot} \bar{A}
\end{align}
and hence
\begin{align}
		A \wedge B \rightarrow A \circ B,
\end{align}
where
\begin{align}
		A \circ B = \frac{1}{2} \left ( A {\cdot} B + B {\cdot} A \right )
		\label{eq:jordanproduct}
\end{align}
is the commutative Jordan product \cite{jordanUberMultiplikationQuantenmechanischer1933}. The ansatz also implies that
\begin{align}
		A \oplus B \rightarrow A {\cdot} \bar{B} {\cdot} A + \bar{A} {\cdot} B {\cdot} \bar{A}.
\end{align}
It can be shown that also this expression is commutative. In sum, due to the L\"uders ansatz all connectives are commutative. Furthermore, the Jordan product also satisfies the marginality relations (\ref{eq:logmarg}). 

Jordan algebras where the Jordan product takes the form (\ref{eq:jordanproduct}) are called \emph{special}. Jordan algebras are a separate research field in mathematics \cite{mccrimmonTasteJordanAlgebras2004}. Although Jordan algebras were originally proposed in physics, their physical meaning has remained obscure until now. It has not been obvious why the observables of a physical system should carry any physically meaningful bilinear product \cite{barnumCompositesCategoriesEuclidean2020}. However, a connection between the requirement of symmetry of the exclusive disjunction and Jordan algebras was noted in Ref. \cite{alfsenNonCommutativeSpectralTheory1979}.

In a logical system, a question being asked twice should give the same answer. Therefore, we require that ${A \sqcap A = A}$. This translates, by (\ref{eq:lueders}) to the requirement $A {\cdot} A {\cdot} A = A$, which is satisfied for $A \cdot A = A$. For the reverse order we have the requirement $B {\cdot} B = B$. It then follows that
\begin{equation}
		A \circ A + B \circ B =  A + B = 0  \quad \Rightarrow \quad A = B = 0.
\end{equation}
This means that the Jordan algebra is \emph{formally real}. Jordan, von Neumann and Wigner \cite{jordanAlgebraicGeneralizationQuantum1934} showed that all special, finite-dimensional, formally real Jordan algebras are direct sums of simple Jordan algebras of three different types: self-adjoint parts of real, complex or quaternionic matrix algebras. These results have also been generalized to Jordan algebras with an infinite number of degrees of freedom \cite{zelmanovJordanDivisionAlgebras1979,zelmanovPrimeJordanAlgebras1983}. Thus, the axiomatic Hilbert space foundations of quantum theory follow from a commutative logic.

A number of recent publications have also shown that formally real Jordan algebras follow from various principles of probabilistic or information theoretical character (see e.g.
\cite{wilceFourHalfAxioms2012,barnumHigherorderInterferenceSinglesystem2014,wilceConjugatesFiltersQuantum2019}).

Quantum theory is usually formulated in terms of a complex Hilbert space. We are not discussing here whether some specific principles single out complex Hilbert space. Discussions on this can be found elsewhere (see e.g. \cite{wilceRoyalRoadQuantum2018}).

\section*{Discussion}

We build the logic on the tautology (\ref{eq:tautology}). It is true for all questions regardless of the answers. Interestingly, some philosophers have taken tautologies as defining the laws of logic \cite{papLawsLogic2002}. However, the possibility that tautologies might redefine classical logical connectives does not seem to have been contemplated.

Since the tautology (\ref{eq:tautology}) and the logical conjunction (\ref{eq:logcon}) apply to questions of any sort, independently of what happens in or between the questions, the logic also applies to generalized measurements as they are currently defined in quantum mechanics. However, in this case logical connectives will not be commutative.

\section*{Acknowledgements}

The author acknowledges valuable comments from Kjetil B{\o}rkje and Leon Loveridge.

\end{document}